%% file: DataAnalysis_ComputingChallenges.tex
\documentclass[11pt,fleqn]{book} 

\usepackage{tcolorbox}
\usepackage{wrapfig}
\usepackage{url}
\usepackage[utf8]{inputenc}
\usepackage{aas_macros}
\usepackage{pdfpages}
\usepackage{multicol}
\usepackage{setspace}
\usepackage{tcolorbox}

\input{structure}

\begin{document}
\pagenumbering{roman}

\includepdf{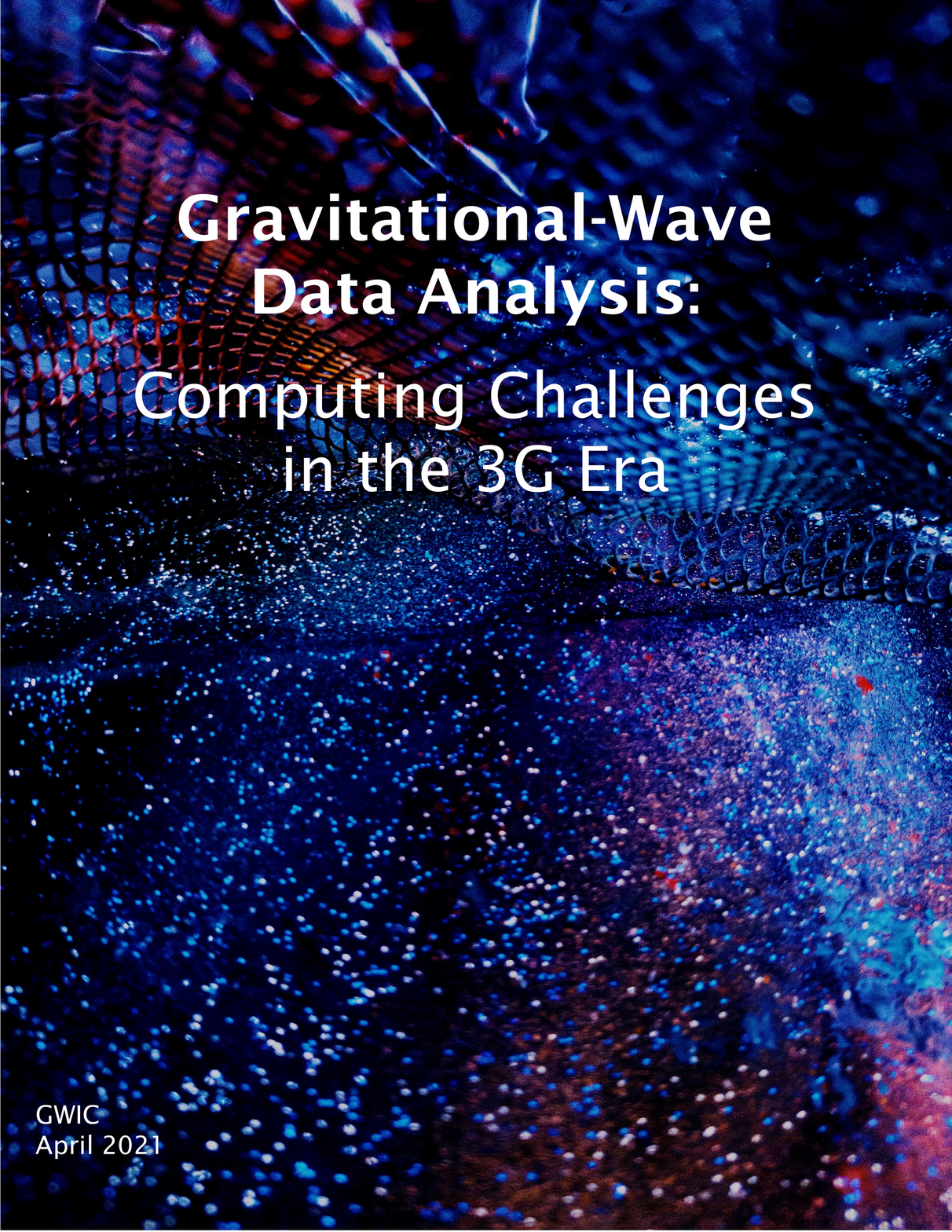}
\input{cover}


\pagenumbering{arabic}
\input{Introduction} 
\input{Challenges} 
\input{Estimates} 
\input{Software} 
\input{Organizations} 
\input{CollabStructure} 
\input{Cyberinfrastructure} 
\input{People} 
\input{International} 
\input{Industry} 
\input{Recommendations}

\clearpage

\clearpage

\bibliographystyle{unsrt}
\chapterimage{bib.jpg} 
\addcontentsline{toc}{chapter}{Bibliography} 
\small
\begin{spacing}{1.00}
\markboth{\sffamily\bfseries Bibliography}{\sffamily\bfseries Bibliography}
\bibliography{thebib}
\vspace{-2cm}
\end{spacing}
\end{document}

%% file: structure.tex

\let\oldenumerate\enumerate
\let\endoldenumerate\endenumerate
\renewenvironment{enumerate}
  {\let\olditem\item
   \renewcommand{\item}{\clearpage\olditem}
   \oldenumerate}
  {\endoldenumerate}

\usepackage[top=2.54cm,bottom=2.54cm,left=2.54cm,right=2.54cm,headsep=10pt,letterpaper]{geometry} 
\usepackage{xcolor} 
\definecolor{ocre}{RGB}{52,177,201} 

\usepackage{avant} 
\usepackage{mathptmx} 

\usepackage{microtype} 
\usepackage[utf8]{inputenc} 
\usepackage[T1]{fontenc} 

\usepackage{titlesec} 

\usepackage{graphicx} 
\graphicspath{{Pictures/}{figures/cosmo/}} 

\usepackage{lipsum} 

\usepackage{tikz} 

\usepackage[english]{babel} 

\usepackage{enumitem} 
\setlist{nolistsep} 

\usepackage{booktabs} 

\usepackage{eso-pic} 

\usepackage{titletoc} 

\contentsmargin{0cm} 
\titlecontents{chapter}[1.25cm] 
{\addvspace{15pt}\Large\sffamily\bfseries} 
{\color{ocre!60}\contentslabel[\LARGE\thecontentslabel]{1.25cm}\color{ocre}} 
{}
{\color{ocre!60}\normalsize\sffamily\bfseries\;\titlerule*[.5pc]{.}\;\thecontentspage} 
\titlecontents{section}[1.25cm] 
{\addvspace{5pt}\sffamily\bfseries} 
{\contentslabel[\thecontentslabel]{1.25cm}} 
{}
{\sffamily\hfill\color{black}\thecontentspage} 
[]
\titlecontents{subsection}[1.25cm] 
{\addvspace{1pt}\sffamily\small} 
{\contentslabel[\thecontentslabel]{1.25cm}} 
{}
{\sffamily\;\titlerule*[.5pc]{.}\;\thecontentspage} 
[]

\titlecontents{lsection}[0em] 
{\footnotesize\sffamily} 
{}
{}
{}

\titlecontents{lsubsection}[.5em] 
{\normalfont\footnotesize\sffamily} 
{}
{}
{}

\usepackage{fancyhdr} 

\fancypagestyle{plain}{\fancyhead{}} 

\makeatletter
\renewcommand{\cleardoublepage}{
\clearpage\ifodd\c@page\else
\hbox{}
\vspace*{\fill}
\thispagestyle{empty}
\fi}

\usepackage{amsmath,amsfonts,amssymb,amsthm} 

\newtheoremstyle{ocrenumbox}
{0pt}
{0pt}
{\normalfont}
{}
{\small\bf\sffamily\color{ocre}}
{\;}
{0.25em}
{\small\sffamily\color{ocre}\thmname{#1}\nobreakspace\thmnumber{\@ifnotempty{#1}{}\@upn{#2}}
\thmnote{\nobreakspace\the\thm@notefont\sffamily\bfseries\color{black}---\nobreakspace#3.}} 

\newtheoremstyle{blacknumex}
{5pt}
{5pt}
{\normalfont}
{} 
{\small\bf\sffamily}
{\;}
{0.25em}
{\small\sffamily{\tiny\ensuremath{\blacksquare}}\nobreakspace\thmname{#1}\nobreakspace\thmnumber{\@ifnotempty{#1}{}\@upn{#2}}
\thmnote{\nobreakspace\the\thm@notefont\sffamily\bfseries---\nobreakspace#3.}}

\newtheoremstyle{blacknumbox} 
{0pt}
{0pt}
{\normalfont}
{}
{\small\bf\sffamily}
{\;}
{0.25em}
{\small\sffamily\thmname{#1}\nobreakspace\thmnumber{\@ifnotempty{#1}{}\@upn{#2}}
\thmnote{\nobreakspace\the\thm@notefont\sffamily\bfseries---\nobreakspace#3.}}

\newtheoremstyle{ocrenum}
{5pt}
{5pt}
{\normalfont}
{}
{\small\bf\sffamily\color{ocre}}
{\;}
{0.25em}
{\small\sffamily\color{ocre}\thmname{#1}\nobreakspace\thmnumber{\@ifnotempty{#1}{}\@upn{#2}}
\thmnote{\nobreakspace\the\thm@notefont\sffamily\bfseries\color{black}---\nobreakspace#3.}} 
\makeatother

\newcounter{dummy}
\numberwithin{dummy}{section}
\theoremstyle{ocrenumbox}
\newtheorem{theoremeT}[dummy]{Theorem}

\newtheorem{exerciseT}{Exercise}[chapter]
\theoremstyle{blacknumex}
\newtheorem{exampleT}{Example}[chapter]
\theoremstyle{blacknumbox}

\newtheorem{definitionT}{Definition}[section]
\newtheorem{corollaryT}[dummy]{Corollary}
\theoremstyle{ocrenum}

\RequirePackage[framemethod=default]{mdframed} 

\newmdenv[skipabove=7pt,
skipbelow=7pt,
backgroundcolor=black!5,
linecolor=ocre,
innerleftmargin=5pt,
innerrightmargin=5pt,
innertopmargin=5pt,
leftmargin=0cm,
rightmargin=0cm,
innerbottommargin=5pt]{tBox}

\newmdenv[skipabove=7pt,
skipbelow=7pt,
rightline=false,
leftline=true,
topline=false,
bottomline=false,
backgroundcolor=ocre!10,
linecolor=ocre,
innerleftmargin=5pt,
innerrightmargin=5pt,
innertopmargin=5pt,
innerbottommargin=5pt,
leftmargin=0cm,
rightmargin=0cm,
linewidth=4pt]{eBox}	

\newmdenv[skipabove=7pt,
skipbelow=7pt,
rightline=false,
leftline=true,
topline=false,
bottomline=false,
linecolor=ocre,
innerleftmargin=5pt,
innerrightmargin=5pt,
innertopmargin=0pt,
leftmargin=0cm,
rightmargin=0cm,
linewidth=4pt,
innerbottommargin=0pt]{dBox}	

\newmdenv[skipabove=7pt,
skipbelow=7pt,
rightline=false,
leftline=true,
topline=false,
bottomline=false,
linecolor=gray,
backgroundcolor=black!5,
innerleftmargin=5pt,
innerrightmargin=5pt,
innertopmargin=5pt,
leftmargin=0cm,
rightmargin=0cm,
linewidth=4pt,
innerbottommargin=5pt]{cBox}

\makeatletter
\renewcommand{\@seccntformat}[1]{\llap{\textcolor{ocre}{\csname the#1\endcsname}\hspace{1em}}}
\renewcommand{\section}{\@startsection{section}{1}{\z@}
{-2ex \@plus -1ex \@minus -.2ex}
{1ex \@plus.1ex }
{\normalfont\large\sffamily\bfseries}}
\renewcommand{\subsection}{\@startsection {subsection}{2}{\z@}
{-2ex \@plus -0.1ex \@minus -.2ex}
{0.5ex \@plus.2ex }
{\normalfont\sffamily\bfseries}}
\renewcommand{\subsubsection}{\@startsection {subsubsection}{3}{\z@}
{-2ex \@plus -0.1ex \@minus -.2ex}
{.2ex \@plus.2ex }
{\normalfont\small\sffamily\bfseries}}
\renewcommand\paragraph{\@startsection{paragraph}{4}{\z@}
{-2ex \@plus-.2ex \@minus .2ex}
{.1ex}
{\normalfont\small\sffamily\bfseries}}

\usepackage[hidelinks,backref=page,pagebackref=true,hyperindex=true,colorlinks=false,breaklinks=true,urlcolor= ocre,bookmarks=true,bookmarksopen=false,pdftitle={Title},pdfauthor={Author}]{hyperref}

\newcommand{\thechapterimage}{}
\newcommand{\chapterimage}[1]{\renewcommand{\thechapterimage}{#1}}

\def\thechapter{\arabic{chapter}}
\def\@makechapterhead#1{
\thispagestyle{empty}
{\centering \normalfont\sffamily
\ifnum \c@secnumdepth >\m@ne
\if@mainmatter
\startcontents
\begin{tikzpicture}[remember picture,overlay]
\node at (current page.north west)
{\begin{tikzpicture}[remember picture,overlay]
\node[anchor=north west,inner sep=0pt] at (0,0) {\includegraphics[width=\paperwidth]{\thechapterimage}};
\draw[anchor=west] (2.46cm,-6cm) node [rounded corners=20pt,fill=ocre!10!white,text opacity=10,draw=ocre,draw opacity=1,line width=1.5pt,fill opacity=.6,inner sep=12pt]{\LARGE\sffamily\bfseries\textcolor{black}{\thechapter. #1\strut\makebox[22cm]{}}};
\end{tikzpicture}};
\end{tikzpicture}}
\par\vspace*{130\p@}
\fi
\fi}

\def\@makeschapterhead#1{
\thispagestyle{empty}
{\centering \normalfont\sffamily
\ifnum \c@secnumdepth >\m@ne
\if@mainmatter
\begin{tikzpicture}[remember picture,overlay]
\node at (current page.north west)
{\begin{tikzpicture}[remember picture,overlay]
\node[anchor=north west,inner sep=0pt] at (0,0) {\includegraphics[width=\paperwidth]{\thechapterimage}};
\draw[anchor=west] (2.46cm,-6cm) node [rounded corners=20pt,fill=ocre!10!white,fill opacity=.6,inner sep=12pt,text opacity=1,draw=ocre,draw opacity=1,line width=1.5pt]{\LARGE\sffamily\bfseries\textcolor{black}{#1\strut\makebox[22cm]{}}};
\end{tikzpicture}};
\end{tikzpicture}}
\par\vspace*{120\p@}
\fi
\fi
}
\makeatother

\usepackage{graphicx}
\graphicspath{{./figures/}}
\usepackage{appendix}
\usepackage{latexsym,amsmath,amsfonts,amssymb,booktabs}
\usepackage[font=small]{caption}
\usepackage{slashed,upgreek,amscd,cancel,tensor,color}
\usepackage{adjustbox}
\usepackage[numbers,sort&compress,square]{natbib}
\usepackage{epsfig,latexsym}
\usepackage{url}
\numberwithin{equation}{section}
\usepackage{doi}
\usepackage{subcaption}
\usepackage{mathtools}
\usepackage{upgreek}
\usepackage{stfloats}
\usepackage{afterpage}
\usepackage{multirow}

%% file: cover.tex

\begingroup
\thispagestyle{empty}



\newpage
\thispagestyle{empty}

\textbf{DATA ANALYSIS COMPUTING CHALLENGES SUBCOMMITTEE}

Peter Couvares, Caltech, USA (Co-chair)

Ian Bird, CERN, Switzerland (Co-chair)

Ed Porter, Universit\'e Paris Diderot, France (Co-chair)

Stefano Bagnasco, INFN, Italy

Geoffrey Lovelace, California State University Fullerton, USA

Josh Willis, Caltech, USA\\

\textbf{STEERING COMMITTEE}

Michele Punturo, INFN Perugia, Italy (Co-chair)

David Reitze, Caltech, USA (Co-chair)

Peter Couvares, Caltech, USA

Stavros Katsanevas, European Gravitational Observatory

Takaaki Kajita, University of Tokyo, Japan

Vicky Kalogera, Northwestern University, USA

Harald Lueck, AEI, Hannover, Germany

David McClelland, Australian National University, Australia

Sheila Rowan, University of Glasgow, UK

Gary Sanders, Caltech, USA

B.S. Sathyaprakash, Penn State University, USA, and Cardiff University, UK

David Shoemaker, MIT, USA (Secretary)

Jo van den Brand, Nikhef, Netherlands
\vspace{1.5cm}

\noindent \textsc{Gravitational Wave International Committee}\\

\noindent {This document was produced by the GWIC 3G Subcommittee and the GWIC 3G Data Analysis Computing Challenges Subcommittee}\\ 

\noindent \textit{Final release, April 2021}\\ 

\noindent \textit{Cover: Andre Moura}


\chapterimage{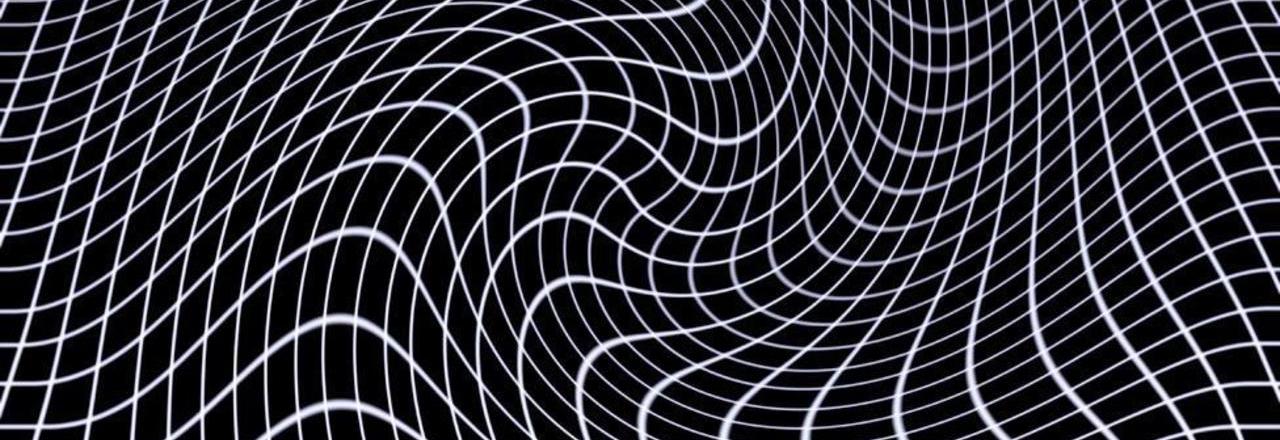} 
\pagestyle{empty} 
\tableofcontents 


\pagestyle{fancy} 
\newpage

%% file: Introduction.tex
\chapterimage{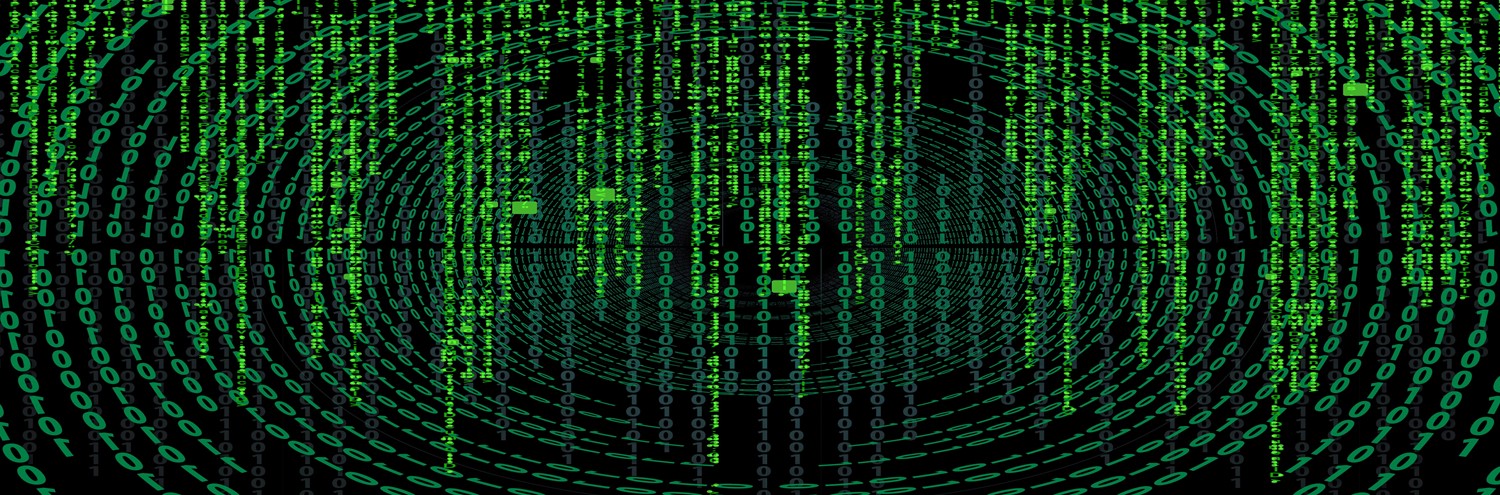} 
\chapter{Introduction}
\label{ch:introduction}

\vspace{0.5cm}
{\Large\bf T}o frame and motivate this report, we begin with background on the existing 2G detector scientific collaborations and an overview of the computing models and methods currently employed. For additional background on the science driving the computational needs, please refer to the 3G Science Case Report.\cite{GWIC3G_SCT}

The Advanced LIGO/Advanced Virgo collaboration (LVC) is composed of three Gravitational Wave (GW) interferometers located in Hanford (WA), Livingston (LA) and Pisa (Italy).  In September 2015, the LVC began a series of advanced era detector runs, with the nomenclature “O\#”.  O1 ran from September 2015 to January 2016, and as well as the first ever detection of GWs, the run ended with the detection of three binary black hole (BBH) mergers.  O2 ran from December 2016 until the end of August 2017.  As well as the detection of a number of other BBH mergers, O2 saw the first ever detection of a merger of two neutron stars (BNS).  O3 began on April 1st 2019, and, due to lockdown imposed by the COVID-19 pandemics, was terminated earlier than scheduled on Mar 27th, 2020.  It is further expected that the Japanese interferometer KAGRA will join the upcoming O4 run.

From a data analysis computing perspective, a primary challenge in the transition from O1 to O2 was the increased computational power needed for both the search and parameter estimation phases.  In the search (detection) phase, the waveform template banks increased in size to accommodate a larger range of masses.  In the parameter estimation phase, while the computational cost of each run remained almost the same as in O1, the greatly increased number of GW sources, as well as the number of exploratory runs that were needed for the BNS merger discovery, caused the computational cost to explode.  In addition, the discoveries provided an opportunity to perform unforeseen and computationally-intensive analyses to measure the Hubble-Lemaitre constant H0, test the validity of GR and to constrain the internal physics of neutron stars.

For its third observing run (O3), the LIGO-Virgo collaboration estimates its ongoing data analysis computing requirements at ~700 million CPU core-hours\footnote{On a reference Intel Xeon E5-2670 2.6Ghz CPU} per year, to perform some 80 astrophysical searches, follow-up activities, and detector characterization activities.  The 10 most computationally intensive of these analyses comprise about 90\% of the demand, with a long tail of the remaining 70.  Most of this computing consists of pleasingly parallel High Throughput Computing (HTC) for “deep” offline searches; ~10\% is for low-latency data analyses needed to generate rapid alerts for multi-messenger (electromagnetic, neutrino) followup.  Very little high-performance parallel computing is required, with the exception of Numerical Relativity simulations, which are not included in this assessment.

While during O1 the vast majority of the computing power was provided by dedicated LIGO-Virgo clusters, either on-site or in large computing centres, during O2 and O3 more and more use was made of external shared computing resources. This growth of shared, external computing resources prompted the development of a distributed computing model, similar to the ones used for example by large LHC collaborations. Furthermore, the Virgo, LIGO and KAGRA collaborations are joining efforts moving from partially interoperable owned computing resources to a wholly shared common computing infrastructure under the IGWN (International Gravitational Wave observatories Network) rubric\cite{Bagnasco2020,WeitzelBBCWH17}, by adopting common tools and interfaces.

The LVC has carefully investigated the use of parallel GPU and MIC architectures for its most compute-intensive searches; CUDA GPUs have been the most successful and cost-effective, and were deployed at scale for the first time in O3.

Currently little use is made of commercial cloud resources for data analysis computing aside from service infrastructure, for example for the low-latency services that generate and distribute the public event candidate alerts. There are no major technical obstacles to performing HTC on the cloud (they are similar to other shared resources), however the logistics and cost-effectiveness of funding and managing metered computing are still being understood, while the use of academic cloud resources (such as the EGI FedCloud in Europe) are being investigated. 

LIGO/Virgo generates $\sim$20TB per interferometer (IFO) per observing year of h(t) strain data used by most analyses, and ~1PB per IFO per observing year of raw data (all channels, full sample rate).  The scale of gravitational-wave data analysis is no longer “big data” by 2020 standards — but data management remains non-trivial in a distributed HTC environment nonetheless.

In the following, we will adopt the following nomenclature:
\begin{itemize}
    \item Analysis pipelines: individual data analysis codes written by domain scientists, on top of Core software libraries and frameworks, to extract the scientific results from data;
    \item Core software: common data analysis applications used by scientists, scientific computing libraries, and experiment software frameworks;
    \item Cyberinfrastructure (US) or e-Infrastructure (EU): services used to run distributed, federated computing, and the corresponding software for data management, workload management, AAI, etc.; Computing facilities: the underlying computing resources (CPU, storage, networks, local support) supporting data analysis computing infrastructure and codes.
\end{itemize}

%% file: Challenges.tex
\chapterimage{Challenges_pic.jpg} 
\chapter{Challenges}
\label{ch:ns}

{\Large\bf T}he demand for data analysis computing in the 3G era will be driven by the high number of detections (up to hundreds per day), the expanded search parameter space (0.1 solar masses to 1000+ solar masses) and the subsequent parameter estimation (PE) follow-up required to extract the optimal science. Given that GW signals may be visible for hours or days in a 3G detector (as opposed to seconds to minutes in the 2G era), their waveforms may exceed physical memory limits on individual computing resources.  On top of this, as the 3G network will be an international effort, there will be an increased need to develop appropriate and scalable computing cyberinfrastructure, including data access and transfer protocols, and storage and management of software tools, that have sustainable development, support and management processes.

Data storage should not be challenging, since GW data products are not expected to grow by many orders of magnitude in the 3G era; the challenge will be in the computational power needed to extract the science content from the data.

In terms of computational challenges for the analysis of gravitational-wave data, there appear two major aspects for the 3G era, relative to 2G: compact binary coalescence (CBC) detection and CBC parameter estimation. 

\section{CBC Detection}
At present, the majority of GW detections are made using a matched-filter based template bank.  This requires covering the parameter search space with many hundreds of thousands of theoretical waveform models (or templates).  The cost of matched-filter searches for CBC signals grows dramatically with improvements in sensitivity as detector bandwidth is increased, so straightforward application of existing methods are unlikely to meet the science requirements of the 3G era.  Specifically, both the number and duration of waveform templates needed increases substantially for the lower frequency limits afforded by the 3G detector designs. Thankfully, the dominant costs are independent of predicted rates and rely on parallel algorithms with independent tasks that are horizontally scalable.  3G detectors will observe vastly more gravitational waves than 2G detectors, including some with tremendous signal-to-noise ratios and some that remain in the detector's sensitive frequency band over much longer times. Given the increased duration and higher number of signals, extracting all of the science encoded in these observations would be prohibitively expensive with today's data analysis techniques; much work will be needed on core software and search pipeline R\&D to reduce the overall computing requirements and overcome new challenges.

Assuming waveform template banks are still applied in the 3G era, a number of questions need to be answered: 
\begin{itemize}
    \item How many templates will be needed to cover 1-1000(+) solar mass events?
    \item Will sub-solar-mass events (0.1-1 solar masses) need to be included in the template bank?
    \item How will matched filtering perform if there are overlapping mergers in the data?
    \item How large must template banks be (i.e., how densely will waveforms need to be placed in parameter space) in order to maintain O2-era SNR targets; or inversely, how much SNR will be lost if template bank size is constrained?
    \item What are the science impacts of massively reducing the number of templates at the cost of sensitivity?
\end{itemize}
A major foreseeable problem is how to answer these questions when the ground-based GW data analysis community is fully occupied with development for 2G science runs. The Gravitational-Wave community is still relatively small, and will need to grow and/or divide its efforts between exploiting the existing 2G data and preparing for future 3G data.

\section{CBC Parameter Estimation}
The cost of CBC Parameter Estimation using Bayesian inference may present an even larger challenge.  Improvements in detector sensitivity translate into orders of magnitude greater detection rates, and PE costs scale linearly with the number of detections.  Furthermore, most PE techniques rely on inherently sequential algorithms (e.g., Monte Carlo Markov Chain, or “MCMC”), whose convergence properties vary from algorithm to algorithm.  To make statements on the parameters of a system, a certain number of statistically independent samples from the posterior distribution are required. The current 2G runtime to achieve this number of samples varies not only from algorithm to algorithm, but also from source to source, with the fastest analysis taking days, and the longest taking weeks. This latency is unacceptable in the 3G era.

The development of new, more convergent (and thus more rapid and efficient) algorithms is a field of research in its own right, with the typical timescale for the development of a new algorithm usually being on the order of a decade. The development of such algorithms for 3G must be an interdisciplinary effort involving GW scientists, experts in statistical inference and professional programmers.  The adaptation of advanced algorithms designed for other domains will almost certainly require extensive modification for GW astronomy applications.  Again, this is something that would take a number of years to accomplish.  Such a development will require a large investment of human effort on top of that required by the development of new detection algorithms.

Also, the scientific requirements for PE in the 3G era are not yet well-defined.  In the 2G era, the LVC performed multiple, computationally-intensive PE analyses on every viable CBC candidate, and yet were constrained by available human effort as much as by computing.  In addition, the computing costs of development, testing, and exploration of PE codes were greater than that of the final PE runs used for publication.

In an era of multiple signals per day, it is unclear if it will be feasible or even desirable to run a deep PE analysis on every CBC candidate. It might be that the science goals demand different degrees of PE investigation (and therefore computing resources) for different candidates.  The answer will obviously be driven by the computing efficiency of PE codes (and the degree of automation possible in the PE process), but the range of scientific scenarios is not obvious due to the many remaining uncertainties: 
what is the lower limit of PE investigation needed to realize our most basic 3G scientific goals, and what is its present computing cost; 
what is the upper limit of useful PE investigation we can currently achieve simply by applying more computing power, given the expected signal quality; 
where between these two limits do we see the maximum scientific benefit as a function of cost; and do we see rapidly diminishing returns for additional computing after a certain degree of precision?  
Are there obvious “tiers” of investigation and cost which we should target for different kinds of candidate signals?  
How should we define those tiers, given the science we hope to extract from the data?  

These questions need to be explored iteratively by GW scientists over the next decade in close collaboration with statistical inference and computing experts, as new computing efficiencies are realized and new scientific goals are identified and translated into PE requirements.

Increasing heterogeneity and complexity of computing platforms drives the need for i) better software engineering and testing; ii) additional organizational expertise and effort in optimization, distributed computing (architecture, engineering, support), and computing management; iii) better tools, services, and processes for sustainable optimization; iv) better education and consulting for scientists/developers who are not first and foremost software engineers; v) automated testing for diverse hardware platforms and environments; and vi) more complex deployment, orchestration, instrumentation, and accounting of DA workflows.

\section{Burst and continuous sources}
Unmodeled burst source searches, a major component of 2G gravitational-wave data analysis computing demand, should scale more or less linearly with the number of interferometers and their observing time; unlike CBC, there is no scaling with detector sensitivity or low-frequency cut-off. Thus, the computational demands of unmodeled burst source detection should be relatively modest in the 3G era.

Evaluating the computing resources needed for Continuous-Wave (CW) signal searches is a challenge in itself, however. In the 2G era, such searches were limited by many orders of magnitude less available computing power than optimal, and thus the increased cost of 3G detection will not pose a fundamentally new problem for CW searches.

All-sky searches, with current pipelines (see, for example, \cite{PhysRevD.90.042002}), scale linearly with both the number of detectors and observation time, but the lower frequency regions do not contribute that much to the computation time, so the required computing power is not expected to explode like in the CBC case. However, aiming for a better intrinsic sensitivity of the searches or for a larger explored parameter space would very quickly increase the computation time.

Targeted searches, looking for signals from known or suspected pulsars, are again linear with both the number of interferometers and the amount of accumulated data, but need a very small fraction of the computing power needed for all-sky searches, even if the number of potential sources will grow.

Ultimately in the 3G era, all-sky searches sensitivity will still be limited by the available computing power (mitigated by algorithm developments and code optimisation), more or less by the same amount in the 2G era given the same scientific objectives.
\clearpage

%% file: Estimates.tex
\chapterimage{ResourceEst_pic.jpg} 

\chapter{Computing Resource Estimates}



{\Large\bf A}pproximately 1 million CPU core-hours of computing were used for parameter estimation of each signal in the LVC’s second observing run (O2), compared to ~10k CBC signals per year expected for 3G IFOs; a 3 orders-of-magnitude increase.

A naive scaling of current matched-filter CBC searches and PE to 3G design sensitivities will require many orders of magnitude more CPU and RAM than 2G.  Moore’s Law alone will not deliver the required performance increase, moreso because it is slowing already due to both cost management by a shrinking set of hardware producers and — barring a breakthrough — the physical limits of silicon transistors.  CERN’s outlook is a 10\%/year performance/cost improvement at constant cost, with very large uncertainties \cite{Misawa2020}.  This does not include the coprocessor/accelerator (GPU, FPGA, etc.) market, but there are no panaceas there; the performance improvements that are being delivered by both increased CPU parallelism and co-processors are becoming more complicated to exploit in software via task and/or data parallelism.  Many existing codes won’t benefit from additional CPU parallelism, and a large amount of effort will need to be dedicated to effectively exploiting such architectures.

Optimistically, more efficient techniques for background estimation may reduce CBC search costs by a factor of a few; machine learning-based searches may live up to their promise in culling signals from non-Gaussian noise.  Even a diminished Moore’s law could help improve the performance of sequential codes by a factor of a few over the next decade.  And newer, more parallel algorithms will track Moore’s Law closer than sequential algorithms.

However, something has to give: barring an unexpected breakthrough, we must prepare to be clever about how and where in parameter space our searches can be less sensitive, in order not to miss what we care about most.  This could be done, e.g., via matched-filtering with sparser template banks and/or less faithful waveforms in the most expensive and/or less scientifically valuable regions of the parameter space, or via reduced-order modeling, waveform compression, greater reliance on unmodeled burst searches, and/or machine learning technologies.  We don’t yet know the right mix of approaches for 3G-scale analysis.  We may need to identify new “sweet spots” of sensitivity vs. cost given our science goals.

If the GW science we can realize will indeed be limited by the efficiency of search and PE codes, targeted investments in data analysis development and optimization today, in advance of the 3G era, may pay outsized scientific dividends. These investments should be driven by first understanding the detection and PE requirements needed for specific discoveries, and by stressing that both the computing infrastructure (for more efficient exploitation of available resources) and core software R\&D and quality management (for pipeline optimisation) are crucial 3G investments.

Even with all the possible innovations in the core software, and the mitigations that can be obtained by refining the scientific strategy, the large number of observations to process will require a fundamentally distributed data analysis system, rather than the dedicated Data Centers mostly used in 2G runs; a transition that already started during O2 and that will take advantage of the large shared cyberinfrastructures for scientific computing that we expect will be available in the 3G timescale.

\clearpage

%% file: Software.tex
\chapterimage{CoreSoftware_pic.jpg} 
\chapter{Core Software (AI, Algorithms, etc.)}


{\Large\bf S}cientific data analysis will necessarily rely on both cyberinfrastructure for distributing and managing those analyses, and the core software and libraries implementing the underlying algorithms. As highlighted in the challenges, and quantified in the resource estimates, the improved sensitivity and corresponding scientific potential of 3G detectors requires that the core software be well-optimized, yet also able to target the diverse hardware of a shared and distributed computing environment.
 
Much of the core software currently used to analyze gravitational waves is written and maintained by domain scientists themselves. Among this are examples of software leveraging GPU \cite{2014PhRvD..90h2004D, 2016CQGra..33u5004U, 2017arXiv170202256G, 2019PhRvD..99h4026W}, CPU vectorization \cite{2008CQGra..25k4029K} and high-performance libraries \cite{2019arXiv190108580S}.  However this individually-driven approach to performance already shows its limitations in second-generation analysis codes. Core software for 3G will require extensive optimization targeting diverse hardware, which needs both software engineering and domain science expertise. In addition to developing this software, careful benchmarking to feed into more detailed estimates of computational costs will be essential. All of these activities, to move beyond the ad-hoc approach of the 2G era, need both coordination and consistent staffing of scientists and developers with the requisite cross-disciplinary expertise.
 
Additionally, the resource estimates of this report are largely based on extending the data analysis techniques used for current detectors to 3G detectors. However, new data analysis techniques are also under constant development, and in particular the success of machine learning in several scientific disciplines has prompted investigations of those methods in gravitational-wave data analysis. We cannot adequately summarize those investigations here; a recent overview is in  \cite{2020arXiv200503745C}. Examples of the applications considered are searches for compact binaries \cite{2018PhRvD..97d4039G, 2018PhLB..778...64G, 2018PhRvL.120n1103G, 2019PhRvD.100f3015G, 2020PhLB..80335330K, 2020arXiv200601509S}, searches for isolated pulsars \cite{2020PhRvD.102b2005D, 2019PhRvD.100d4009D}, parameter estimation for compact binaries \cite{2020arXiv200207656G, 2019arXiv190906296G, 2020PhRvL.124d1102C}, and detector characterization \cite{2018PhRvD..97j1501G, 2017CQGra..34f4003Z, 2018CQGra..35i5016R}.
 
Some of the papers above find machine learning techniques to be extremely promising; others find those techniques to compare poorly to traditional approaches. More relevant will be the state of the art in the 3G era. Some of the most successful applications of machine learning---such as long short-term memory for speech recognition \cite{ NIPS1996_1215, Hochreiter1997LongSM}, or convolutional neural networks for image classification \cite{6795724, Cun90handwrittendigit}---come from machine learning techniques that are adapted to particular problems. It is not unreasonable to expect that the full potential of machine learning to gravitational-wave data analysis will likewise benefit from similar domain-specific techniques.
 
Both hardware-targeted optimization of traditional techniques, and production-ready machine learning techniques, will need to be tested across realistic data sets. The most appropriate technique for each data analysis task, whether search or parameter estimation, will also depend on the time frame in which results are needed: “early warning” pre-merger detection, low-latency CBC detection and follow-up, or higher-precision but higher-latency “deep” analyses. Optimal hardware procurement will also need to be carefully benchmarked on validated and tuned production-ready core software.
 
Taken together, all of these considerations emphasize the need for both a stable organizational structure in which cross-disciplinary expertise can be coordinated, and formal mock-data challenges that rigorously compare different approaches, similar to those successfully used by LISA.  A LISA challenge begins with publicly released simulated noise, with simulated signals hidden inside. Anyone in the community is then welcome to build and apply computational tools that attempt to find and interpret the hidden signals. Such activities provide increasingly precise estimates of required computing resources, and inform scientific prioritization if resources must be constrained. As an example, the data challenges spurred the development of stochastic search techniques, because a template-bank approach would not be computationally feasible. In the 3G case, mock-data challenges can also be used to test the readiness and usability of the emerging distributed computing cyberinfrastructure, similarly to what the LHC experiments did in the early years of what later became the WLCG.

Recommendation: The community should organize mock-data challenges for 3G data analysis, similar to ones undertaken by the LISA community, to systematically test competing techniques and ensure that the best approaches are robust and production-ready.

\clearpage

%% file: Organizations.tex
\chapterimage{Organizations_pic.jpg} 
\chapter{Organizations}


{\Large\bf T}he GW community is rapidly moving from an era of single experiments managing their own computing needs, to a time of global collaboration between GW experiments, within a more and more complex international environment where it will be competing for resources with other sciences that are also exploring very large scale data sets, and other global collaborations.  As it stands today the GW community does not have a mechanism for organizing a strong collaboration in computing.  We believe that in preparation for the future it is timely to consider what a global computing collaboration for GW would look like and the advantages that would arise from having such a structure.

It will be important for the GW community to have a strong and recognised voice with the cyberinfrastructure communities who are advancing collaborative, federated, global computing services.  Thus it will be very important to create a GW computing management structure to have a clear input in discussing and collaborating with other large research computing efforts as well as the cyberinfrastructure.  This organization must gather and represent the needs of the global GW community, and to be able to negotiate and plan with a clear, recognizable voice.  There are several reasons behind this:
\begin{itemize}
    \item The GW community has challenges in terms of the nature of its workflows and resource needs, that may not be well represented by the existing cyberinfrastructure communities (for example large memory requirements, use of complex pipelines, specific CPU/GPU needs).
    \item In an era where GW does not expect to have owned, dedicated resources, a body is needed to address common challenges and explore synergies with others.  Computing for GW will be co-located in data centres used by other large-scale sciences (LHC, SKA, etc.)
    \item Similarly, a body is needed to address other cyber infrastructure needs on international resources: facilities, networks, etc., either in collaboration with other sciences or in representing GW-specific needs.
    \item It is needed as a single political voice to face funding agencies, resources providers, etc.
    \begin{itemize}
        \item This is essential to have to argue coherently internationally for people and hardware resources.
        \item It represents a point of organization to collaborate on project proposals - and to be coherent between those across different Funding Agencies.
        \item This is where the WLCG (in the case of High Energy Physics, HEP) has been very successful - the funding agencies recognise the global nature and organization and are open to discussion on how to collaborate on resources. 
    \end{itemize}
    \item A management body is needed to agree on responsibilities between various GW collaborations; at various levels - perhaps MOUs for services, data and other policies, resource provision, etc.  This will include GW presence in open data policy and funding, open access activities and would be coherent with that in other Astronomy of physics domains.
    \item A global computing collaboration management body could also oversee and commission community-wide technical developments needed for progress, and to help integrate with the global infrastructures it wants to use; for example on Authentication and Authorization Infrastructure, or other areas.  Such a body would also be able to oversee common operational support (across GW collaborations) in such key areas as security coordination, in conjunction with other cyberinfrastructures and so on.  It can provide a presence in relevant technical bodies (e.g. those of EOSC, OSG, ASCR, PRACE, etc.).  Again, this organization can work internationally to make best use of all of the funding opportunities available, and to be coherent across them.
\end{itemize}

A global computing management could also address specific needs where good coordination is required:
\begin{itemize}
    \item There is a management challenge in balancing between effort needed for support of the current observing runs and the effort needed for future developments and preparation.  This could be addressed with a global view of effort across the collaborations.
    \item Given the length of time to develop new algorithms, the management body could also set up groups to tackle specific problems with experts drawn from across the experiments, and help to resource and fund these.
    \item Similarly working groups on “R\&D” topics may be required for new search algorithms, PE, use of new computing architectures, etc. Again a common management will be key.
\end{itemize}

%% file: CollabStructure.tex
\chapterimage{CollabStructure_pic.jpg} 
\chapter{Suggested Collaboration Structure}


\vspace{3mm}
{\Large\bf D}rawing lessons from other sciences operating at a global scale, we propose some strategies for management that seem to work well.

First, while it is essential that such a global management structure have the support of the experiment/detector/observatory managements, it should also not be too prescriptive and must have the buy-in of the technical teams.  In the case of LHC and the WLCG, this mandate comes via a formal MoU between the experiments, the facilities, and the funding agencies.  This MoU covers both the contributions of computing and storage resources, but also the agreements of how the collaboration will work, respective roles and responsibilities.  In the LHC case this is a very formal MoU but implemented in a collaborative and non-confrontational way.

A simple management body (here referred to as a Management Board - MB) would oversee ongoing work and organise collaborative developments etc.  Its members would be drawn from all of the stakeholders, the GW collaborations, the cyber infrastructures, and the computing facilities that are funded to provide resources to GW.  It would call upon specific technical expertise as needed.  It would meet regularly (e.g., monthly), and manage the ongoing operations, set up working groups for development, and provide the voice of the global GW collaboration in computing.  The bullet points enumerated above would be its key tasks.

Such a body must report back to the stakeholders once or twice a year, for example.  It should have scientific review from the experiments, but also report back to the funding agencies in some form.  In this way, trust can be built, and new requests for resources will have a foundation that is already understood. It is also important to recognise that this feedback and review ensures that national interests are served. This body could in turn establish technical working groups that would report to it, focused on specific developments, software, operations, infrastructure, etc.

Once such a management structure is in place, a medium and long term plan for resource and funding needs should be constructed.  This plan would be updated regularly as part of the stakeholder review, and while of course it will evolve and have many uncertainties, it will nevertheless provide guidance to the funding agencies and other resource providers.  This long-term planning aspect is one of the important benefits of an organisation as suggested here.

This organization and initial plan should be put in place as early as possible in order to get buy-in, and if it is decided to have a formal agreement, the MoU process will take some time to implement. At the same time as this organisation is being set up, a first computing roadmap should be written, initially as a discussion document, but then leading into detailed technical plans. These initial steps will provide strong leadership in the computing activities for GW, and provide a foundation for long-term collaborative investment.

\vspace{3mm}
\textbf{Recommendation:}
\begin{itemize}
    \item Set up now a computing collaboration for GW, with a structure of a management body and appropriate stakeholder review as discussed above.  The collaboration could be bound by an MoU structure.
    \item Provide an initial resource plan (hardware and effort) with an outlook of a few years, to be updated yearly, and maintained as the current best outlook.
    \item Provide a computing technical strategy and roadmap, for discussion and later to become implementation plans.
\end{itemize}

%% file: Cyberinfrastructure.tex
\chapterimage{Cyberinfrastructure_pic.jpg} 
\chapter{Cyberinfrastructure, Resources, Facilities}


\vspace{3mm}
{\Large\bf G}iven that the hardware resources for 3G will be orders of magnitude larger than 2G, how will these resources need to be organized and managed?  A centralized “walled garden” of homogeneous, dedicated GW clusters likely to be inefficient and prohibitively expensive.  A diverse worldwide GW collaboration is unlikely to select a single external computing provider (commercial or otherwise).  3G computing optimization may require specialized hardware for different searches.  The landscape of 3G computing resources will thus necessarily be both more heterogeneous and more distributed than that of the 2G era.

The larger organization of the GW community will also have a profound effect on the organization of computing.  A highly-centralized collaboration with proprietary data requires and enables a more centralized computing organization, whereas a decentralized community of scientists with open data necessitates more loosely-coupled forms of organization.  The implications of the larger GW community structure on computing management and infrastructure must be taken into account when considering the pros and cons of competing visions for future 3G GW communities.

An improved cyberinfrastructure will be needed to make a more complex, dynamic network of data analysis resources usable and robust for a distributed collaboration of scientists.  This includes distributed scheduling, identity management and access control infrastructure, cybersecurity tools and services, data analysis software development and testing tools, resource accounting and reporting tools, low-latency alert and coordination infrastructure, as well as public data and code release infrastructure.

We expect increasing heterogeneity and complexity of computing platforms in the 3G era:
\begin{itemize}
    \item of processing hardware; due to the opportunities for cost savings, data analyses may need to support multiple generations of CPUs, GPUs, MIC platforms and treat them each as distinct platforms. “Lowest common denominator” code capable of running on any platform may not be efficient enough.  
    \item of providers — resources will include those internal to the project, partners \& collaborators, institutional, regional/national, commercial, volunteer.
    \item of target operating systems and software environments. Containerization, etc. are tools to help mitigate this complexity but carry their own costs and are not a complete solution.
    \item of batch/queueing systems.
    \item of storage and network interfaces and capabilities.
    \item of policies for identity and access management, workflow prioritization.
    \item of accounting models and accounting systems
    \item of motivations and expectations — e.g., mutual scientific or strategic interest, public or scientific recognition, financial or other compensation, etc. — not all of which will be explicitly spelled out in MOUs, SLAs, or contracts.
\end{itemize}

%% file: People.tex
\chapterimage{PeopleResources_pic.jpg} 
\chapter{People Resources}


\vspace{-.5cm}
{\Large\bf T}wo major challenges to computing in the 2G era have been uncertain and discontinuous funding streams for computing labor embedded within the collaboration groups outside the IFO laboratories, and a mismatch between existing job categories and new hybrid scientific computing roles.  The need to professionalize software development and engineering, and support increasingly complex computing environments, demands more full-time professional computing expertise in side-by-side collaboration with domain scientists (as opposed to part-time volunteer/service work by scientists, or non-collaborative computing service providers).

LIGO’s recent experience has shown that the return-on-investment of dedicated, “full-stack” computing optimization by a team of both GW science and computing experts is overwhelmingly positive, in terms of saved computing costs for expensive data analysis codes.  The GW community should plan on making this investment for 3G.  Data analysis efficiency improvements needed for 3G will require increased software development effort.  This will require time from busy GW scientists and/or the hiring of more computing specialists.  An investment in GW-fluent computing specialists can free up GW scientists to focus on GW science.

Long-term software development costs may also be higher for rapidly-evolving parallel hardware platforms (GPU, MIC, AVX512, FPGA, etc.) than for traditional CPUs, given that parallel programming interfaces are less stable targets.  Most single-threaded CPU codes have run on new hardware with minimal modifications for more than 30 years.  Data analysis on more distributed, non-dedicated computing grid/cloud platforms will require computing infrastructure investments.  This will also require time from busy GW scientists (serving as a new “tax” on their research output) or the hiring of more computing specialists.

In total, 3G computing labor costs must go up by a factor of a few relative to existing 2G computing models.  Notably, there may be potential and interest in broader collaboration in this area, e.g. with HL-LHC (and other HEP like DUNE) as well as SKA. Many of these are not traditional software development or IT roles that can be outsourced beyond the project — they are hybrids of domain science, research computing, consulting, software engineering, and distributed systems development and operations roles.  These roles benefit enormously from institutional and project “memory” — large, distributed scientific projects pay dearly in time, money and quality when domain-specific experience and relationships are lost.

It is also difficult to recruit and retain career professionals on overlapping 1-3 year awards, and to find research funding for computing, which is not always “transformative” science in and of itself, but needed to enable transformative science.  This is an old problem but will become more acute with the increasing need for these computing roles in the 3G era.  Recognition and job security at the employer level (at universities and institutes) will be critical.

These concerns are common within other science communities, and could help to drive a common effort to recognise software and computing as scientific career tracks, as well as collaborate in a common lobby for more focused investment in software.

\vspace{3mm}
\textbf{Recommendation:} In concert with HEP and Astronomy communities, we recommend that the GW community strongly advocate to funding agencies to provide more stable funding streams and, along with university and institute employers, define and recognize new roles for hybrid domain science / Computer Science staff dedicated to the computing challenges discussed in this report.

\vspace{2mm}
\textbf{Recommendation:} Long-term career opportunities must be provided for GW-fluent computing professionals to support domain scientists; the bulk of computing infrastructure and core software development effort cannot be effectively delivered in the 3G era by domain scientists or external (non-GW-fluent) computing professionals.

%% file: International.tex
\chapterimage{InternationalCollab_pic.jpg} 
\chapter{Global Collaboration, Other Sciences, Common Challenges}


{\Large\bf A}s the GW community prepares for the next generation of detectors and associated computing challenges, it is opportune that several other sciences in Astronomy and HEP are facing similar challenges.  These include managing large scale global computing and data distribution, co-existing in scientific data centres with other sciences, and planning the evolution of computing infrastructures and software in order to benefit from evolving technology, whilst maintaining stability and reliability.  In addition, new challenges of open data, data sharing and collaboration, and other policy-level concerns are coming to the forefront.

There are several dimensions to the potential for collaboration, with a number of opportunities:
\begin{itemize}
    \item Collaboration between scientific collaborations on basic technology concerns: evolution of processors and the difficulty of obtaining the best performance, how to manage this in the long term with an outlook of continually changing hardware.  This area includes also how to manage the change of technology (e.g. if tape were to disappear); the unpredictability of cost evolution of computing and storage, etc.; use of novel architectures and systems (HPC, HTC, etc.).  
    \item Common software challenges, often in order to manage the previous point; here opportunities are many, with the HPC community internationally, and also with entities like the HSF, and software institutes in the USA and EU.
    \item Co-location of GW experiments in data centres together with other sciences.  This can be very costly for the support in those centres unless there are some commonalities and synergy between the resources and services required by all of the experiments supported.  In some cases the centre may mandate that the experiment use a certain type of resource.
    \item Collaboration with other Astronomy (CTA, Vera Rubin Observatory, SKA, etc.) and similar particle physics collaborations.  Engagement across the board to explore synergies and common interests is useful and potentially hugely beneficial in driving towards common solutions.
    \item Collaboration with international cyberinfrastructures (OSG, EGI, EOSC, etc.) and how to benefit from the resources and services they provide, but also how to benefit from the funding opportunities that may come via those infrastructures.  Often such funding will be expecting broad synergies and will not fund a specific experiment alone.  Also such cyberinfrastructures can provide key collaboration points on policy, cyber security (policy and reaction), agreement on global coherence of federated Authentication and Authorization Infrastructure systems, and so on.
    \item Policy development and interaction with funding agencies from a broad scientific consensus.  Common statements from a wide range of (similar) sciences (e.g. astronomy, astroparticle, GW, HEP, NP) could have a very big impact with funding agencies on common concerns.  This would also help in a common strategy for provision of network bandwidth and capability, and areas such as large scale use of HPC, which may require a different access and usage policy for our communities compared to their traditional user base.
    \item Opportunity to engage with global collaboration on scientific computing infrastructure (see below).\\
\end{itemize}

In many of the above, bottom-up collaboration on topics of common interest is often the most direct and beneficial.  However, a management view is needed to ensure coherence and to avoid duplicate or competing efforts, especially with limited resources. Thus the GW community would be well advised to work together with their scientific and international peers in order to exploit such opportunities.  The management organisation proposed in an earlier section would take the lead on organising this.

The HEP community and others, has a working group looking into the evolution of the cost of hardware (CPU, disk, tape) in preparation for the HL-LHC upgrades on the 2026 timescale.   Lessons from that study are relevant also to GW. CERN and other institutes have tracked costs over more than 15 years, and use recent trends to set expectations for the evolution over the coming years.  Simple “Moore’s law” predictions no longer seem to be valid and are outweighed by market forces and cost management by the relatively few hardware producers.  It is a fact that there are now only a handful of chip fabrication companies, only two HDD manufacturers and that tape drive technology is a monopoly with only IBM building drives and only two media producers.

For the past several years the HEP community has used as guidance a 20\% per year improvement in performance for CPU and disk storage capacity for a constant cost, to predict the affordable capacity of complete server systems (not the raw disk space or raw CPU performance).  For CPU this applies to standard x86-like processors.  For tape storage capacity the capacity increase per year was estimated to be somewhat more, around 25\%/yr.

In the last 2 years however we have observed a worrying trend, that there are large fluctuations in the market prices of key components (in particular RAM), and shortages of supply, and that the effective growth is now only around 10\% per year for constant cost.  In addition all of these markets are now in the hands of very few large manufacturers, who control the markets, and thus prices.  The upshot of this is that it is currently very difficult to make a realistic estimate of the price/performance evolution for the coming years.  This is a significant concern when trying to understand the overall cost of computing and storage for future experiments.

\vspace{3mm}
\textbf{Recommendation:} Identify (and if necessary, establish) the right forums for the GW community, HEP, and Astro communities to regularly and systematically discuss the overlap in our computing problems, share information, and plan specific technical collaborations where useful. For example:
\begin{itemize}
    \item Coordination of Distributed Scheduling, Computing Security, and Identity and Access Management efforts between HL-LHC and LIGO/Virgo.
    \item Computing optimization approaches, techniques and tools.
    \item Software engineering tools and technologies.
\end{itemize}

Some of this interaction is already happening (e.g., the ESCAPE project), but the GW community should expand those efforts and have a stronger voice in forums where these matters are discussed.  The 3G gravitational-wave community cannot afford to reinvent wheels.

The high-luminosity LHC experiments face similarly-daunting computational scaling problems over the same timeframe, but they have an order of magnitude larger starting requirements — they are helping to blaze a trail in optimization and distributed computing infrastructure which we can and should follow and collaborate on.  Being explored are Exabyte-scale capabilities for data management, able to serve data in an efficient way to heterogeneous and globally distributed compute resources including HTC clusters, HPC clusters, and commercial clouds.  We foresee the requirement to support a variety of processors ranging from CPU, accelerators (e.g., GPU, MIC), FPGA and other innovative architectures.

CERN/WLCG is basing its strategy on 3 areas of investment:
\begin{itemize}
    \item A federated data cloud (“data lake”) capable of the long term curation and serving of exabyte-scale data, complemented with tools and services to allow experiments to manage that data and content delivery tools to efficiently serve the data to remote compute resources.  This data lake would also provide an effective mechanism for hosting and serving data to the broader community including public open access, potentially with the inclusion of commercial interests.
    \item A significant investment in application software skills.  The ability to make use of the full set of available resources in coming years requires development of software skills and the ability to move applications to appropriate compute architectures in an agile way as they evolve.  The infrastructure will also depend on a clearinghouse of software tools and services for science communities.  The HEP community has initiated the HEP Software Foundation (HSF) as the vehicle through which to address the long term software investment and also to provide the locus for a software tools clearinghouse.  It is not necessarily specific to HEP, and could also be strengthened with the inclusion of other science communities.
    \item Networks to support the infrastructure and the use of appropriate technologies to manage data management both within the data lake and in serving data externally.  WLCG has developed an overlay network (LHCONE) that the National Research and Education Networks use to manage the science data traffic.  This concept has proven extremely versatile and is now used by many more science communities.
\end{itemize}

In addition to the above infrastructure developments, CERN has pioneered aspects of open access and open data, including the Zenodo publishing platform and the CERN open data portal, as well as the storage infrastructure underpinning these.  These could be interesting components of a future service for the GW community.  CERN also provides “CERNBox” - a DropBox-like open source solution that is widely used in the HEP community.  

To complete the picture, CERN has the “openlab” concept, which is a vehicle for collaboration with industry, disconnected from any expectation of procurement.  This provides a mechanism to fund and staff projects of joint interest between CERN and various industrial partners.  In recent years this has expanded to include other research institutes.  This is another avenue of potential technology transfer to the 3G GW community, enabling the investigation of new computing technologies in collaboration with the companies, and to provide access to their expertise.

As noted in our discussion of core software, optimal performance of that software will depend on carefully targeting a diversity of hardware, and hardware acquisition should be based onoff of systematic benchmarking. Achieving both of these goals can be accelerated by careful involvement of industry partners. Though the success of CERN’s openlab is mixed, it provides a useful starting point for a gravitational-wave model. Ideally, such a collaboration will allow early access to testing platforms to inform purchase decisions, training of scientists through fellowships or internships, and engagement on industry-supported coding platforms and software development tools.

From the CERN viewpoint collaborating with the GW community on some or all of these aspects and the related technologies is of great interest and could aid in developing a broad toolset useful to a broad scientific base.  In the recent European Strategy for Particle Physics meeting (May 2019), a broad collaboration of HEP with the Gravitational Wave community was seen as strategic, and very much encouraged from both communities.  

%% file: Industry.tex
\chapterimage{Collab_w_Industry_pic.jpg} 
\chapter{Collaborations with Industry}


{\Large\bf C}ollaboration with industry partners can be beneficial in certain areas. The CERN openlab experience was that, in R\&D topics, such a collaboration can be extremely useful to both partners, and it is also a very good means for training and engaging young scientists and computing/software engineers, and for funding summer students. It should be recognised that management of such projects and people also potentially takes effort away from core tasks. Thus it must be clear what the real benefits and outcomes of such collaboration will be. Experience with CERN openlab shows that an institute would be the point of contact with the potential industry partners to manage all of the legal and IP concerns. The GW community would need to channel this type of engagement through one or two large institutes, or potentially collaborate with CERN on it.

Use of industrial partnerships to obtain services is potentially a risk. We must consider long-term sustainability of the computing infrastructure.  Negotiating a beneficial cost for a service, which then requires specific adaptation may not be a good choice in terms of sustainability, as when the need from our side changes, or the industry partner policies change, the solution may no longer be cost effective, and requires significant effort to replace.  There are many examples of this.  Also this type of partnership brings very little benefit to the science community, and risks embedding proprietary solutions.

\vspace{3mm}
\textbf{Recommendation:} Identify a forum, in collaboration with other Astronomy and HEP organisations, as a means by which to work effectively with industry partners on well identified key projects.

%% file: Recommendations.tex
\chapterimage{SummRecommendations_pic.jpg} 
\chapter{Recommendations}


{\Large\bf T}he GW community is facing a number of challenges as it moves towards the third generation of experiments.  It is clear that a global collaboration between the various experiments will be essential, and that collaboration must apply to the core software and computing activities as well, or the full scientific potential of the instruments will not be realized.

Resources, both physical and human, will be limited, and efforts must be taken to organise the community to be able to use current experience to find new algorithms and computing solutions, exploit synergies, and find new cost efficient workflows, not just between the various GW collaborations, but also with other sciences (Astronomy, Astro-particle, Particle Physics) that are dealing with similar problems, and relying on very similar infrastructures and resources to address them.

In order to do this effectively there are a number of high level recommendations that need to be addressed in order to prepare for the future.  These are the following:

\vspace{3mm}
\begin{itemize}
    \item \textbf{Recommendation:} Set up now a computing collaboration for GW, with a structure of a management body and appropriate stakeholder review as discussed above.  The collaboration could be bound by an MoU structure.
\vspace{3mm}
    \item \textbf{Recommendation:} Provide an initial resource plan (hardware and effort) with an outlook of a few years, to be updated yearly, and maintained as the current best outlook.
\vspace{3mm}
    \item \textbf{Recommendation:} Provide a computing technical strategy and roadmap, for discussion and later to become implementation plans.
\vspace{3mm}
    \item \textbf{Recommendation:} Organize mock-data challenges for 3G data analysis to systematically test competing techniques and ensure that the best approaches are robust and production-ready.
\vspace{3mm}
    \item \textbf{Recommendation:} In concert with HEP and Astronomy communities, we recommend that funding agencies provide more stable funding streams and, along with university and institute employers, define and recognize new roles for hybrid domain science / Computer Science staff dedicated to the computing challenges discussed in this report.
\vspace{3mm}
    \item \textbf{Recommendation:} Identify (and if necessary, establish) the right forums for the GW community, HEP, and Astro communities to regularly and systematically discuss the overlap in our computing problems, share information, and plan specific technical collaborations where useful.
\vspace{3mm}
    \item \textbf{Recommendation:} Identify a forum, in collaboration with other HEP and Astronomy communities, as an appropriate means by which to work with industry partners on well identified key projects.\\
\end{itemize}
\textcolor{white}{white text here white text here white text here white text here white text here white text here white text here white text here white text here white text here white text here white text here white text here white text here} 